\documentstyle[aps,prl,epsfig,amssymb,balanced,times]{revtex}
\begin{document}
\title{Extensive Chaos in the Nikolaevskii Model}
\author{Hao-wen Xi$^1$, Ra\'ul Toral$^2$, J. D. Gunton$^3$ and
Michael I. Tribelsky$^4$}
\address{$^1$Department of Physics and Astronomy,
Bowling Green State University, Bowling Green OH 43403\\
$^2$Instituto Mediterr\'aneo de Estudios Avanzados (IMEDEA),
CSIC-UIB, E-07071 Palma de Mallorca, Spain\\ $^3$Department of
Physics, Lehigh University, Bethlehem, PA 18015\\ $^4$Department of
Applied Physics, Faculty of Engineering, Fukui University, Bunkyo
3-9-1, Fukui 910-8507, Japan\\}
\maketitle
\begin{abstract}
We carry out a systematic study of a novel type of chaos at onset
(``soft-mode turbulence") based on numerical integration of the
simplest one dimensional model. The chaos is characterized by a
smooth interplay of different spatial scales, with defect
generation being unimportant. The Lyapunov exponents are calculated
for several system sizes for fixed values of the control parameter
$\epsilon$. The Lyapunov dimension and the Kolmogorov-Sinai entropy
are calculated and both shown to exhibit extensive and
microextensive scaling. The distribution functional is shown to
satisfy Gaussian statistics at small wavenumbers and small
frequency.\\
\end{abstract}

\begin{twocolumns}
Spatiotemporal chaos (STC) is a subject of considerable
experimental and theoretical importance and occurs in a wide
variety of driven, dissipative systems\cite{CH93,PM90,HG98}. Such
chaotic behavior in spatially extended systems is extremely
difficult to characterize quantitatively, as the dynamics involves
a large number of degrees of freedom. The most common and useful
tool for the characterization of chaos is given by the Lyapunov
exponents $\{\lambda_i\}$. Knowledge of this Lyapunov spectrum
permits one to estimate the number of effective degrees of freedom
of the system (i.e., the dimension of the attractor), using for
example the Kaplan-Yorke\cite{KY79} formula for the Lyapunov
dimension $D(L)$, where $L$ is the linear system size. It also
permits one to test the important concept of extensivity of chaos,
defined as the case in which $\lim_{L\rightarrow\infty} D(L)\sim
L^{d}$, where $d$ is the spatial dimension of the
system\cite{CH93,HS89}. An interpretation of extensive chaos is
that the whole system can then be thought of in some sense as the
union of almost independent subsystems. This was originally
proposed by Ruelle\cite{DR82}, who argued that widely separated
subsystems of a turbulent system should be weakly correlated, so
that the spectrum of Lyapunov exponents would be the union of
exponents associated with each of the subsystems. The question is
closely related to the fundamental problem of ergodicity of
nonequilibrium systems. If the chaos is extensive and each
subsystem evolves in time practically independently of the others,
then in a steady (non-transient) chaotic state the time average is
equivalent to the ensemble average and the system should be
ergodic. Much work has focused on attempting to characterize
spatiotemporal dynamics in these terms (see, e.g.,
\cite{VY89,LLPP93,CH95,WH99,COB99}). However, in spite of the
fundamental importance of the question practically all the results
are related just to a few discrete coupled map
lattices\cite{LPR92,HC93,PP98} and two continuous systems, namely
the complex Ginzburg-Landau (CGL) and Kuramoto-Sivashinsky (KS)
equations (see, e.g., refs.\cite{HC93,PM85,EG95}). This is partly
due to the computational complexity of the problem but primarily to
the lack of simple models exhibiting the requisite chaotic
behavior. It would therefore be of considerable interest to
characterize quantitatively other types of STC.

Recently attention was drawn to the existence of a new wide class
of systems displaying such a behavior\cite{RBR95,KHH96,TT96,MIT97}.
Their properties are qualitatively different from those of the CGL
and KS models. In contrast to both these models the chaos is
associated with smooth, random long-wavelength modulations of a
short-wavelength pattern, with defect generation being unimportant.
The short-wavelength pattern arises due to a single supercritical
bifurcation of the Turing type such as occurs in
Rayleigh-B\'{e}nard convection. The long-wavelength modes belong to
a Goldstone branch of the spectrum originated in a broken
continuous symmetry. The symmetry makes the system degenerate to
the extent that instead of a single, unique spatially uniform
state, it has a {\it continuous family} of equivalent spatially
uniform states, which may be obtained from each other by the
symmetry transformation. This symmetry, which is additional to the
trivial groups of translations and rotations, can be one of many
different types. For this reason the STC in question is quite a
common phenomenon and occurs, for example, in electroconvection in
liquid crystals\cite{RBR95,KHH96}, in convection in a fluid with
stress-free boundary conditions\cite{SZ82,BB84,XLG97,N}, etc., see
ref.\cite{MIT97} for further discussion. The chaos observed in such
cases may be interpreted as a macroscopic dynamical analog of
second order phase transitions, where the order parameter is
related to the amplitudes of turbulent modes. Due to this analogy
it has been called {\it soft-mode turbulence} (SMT)\cite{KHH96}.
The simplest model exhibiting SMT was introduced by
Nikolaevskii\cite{VN89,BN93} to describe longitudinal seismic waves
in viscoelastic media. In what follows we exploit the simplicity of
this model to shed light on general features of this new type of
STC.

We present a detailed systematic study of the Nikolaevskii model,
including calculation of the Lyapunov exponents, fractal dimension,
etc. We show that the Lyapunov dimension and Kolmogorov-Sinai
entropy are extensive quantities, which supports the validity of
the Ruelle's ideas for SMT. We also show that the power spectrum
can be described by Gaussian statistics for small wavenumbers and
frequencies, in agreement with a general argument of Hohenberg and
Shraiman\cite{HS89}.

The model is defined by the following partial differential equation
for the real scalar field $v(x,t)$ (longitudinal mode of the
displacement velocity in the original formulation)\cite{VN89,BN93}:
\begin{equation}
\frac{\partial v}{\partial t} + \frac{\partial^{2}}{\partial x^{2}}
[\epsilon - (1 + \frac{\partial^{2}}{\partial x^{2}})^{2}]v +
v\frac{\partial v}{\partial x}= 0
\end{equation}
with $0\leq x \leq L$ and periodic boundary conditions. This model
has two control parameters, $\epsilon$ (the distance from onset)
and $L$, in contrast to, e.g., the KS model, where the only
non-trivial control parameter is the system size $L$. An essential
feature of the model is that even at small $\epsilon$ it cannot be
reduced to any $\epsilon$-free form\cite{MIT97,BM92}, which makes
the hierarchy of characteristic scales much more complex than those
in the KS and CGL models\cite{TV96,MT97}. Eq. (1) may be regarded
as a generalized Burgers equation and shares with it the same group
of symmetry, namely trivial symmetries under shifts of the
spatiotemporal coordinate system, and nontrivial invariance with
respect to the Galilean transformation $v(x,t)\rightarrow
v(x-v_{o}t,t)+v_{o}$, where $v_{o}$ is an arbitrary constant. The
Galilean invariance plays the role of the above specified
additional symmetry, generating for Eq. (1) the continuous family
of solutions $v=v_{o}$. The Nikolaevskii equation admits a
continuous set of spatially periodic, stationary solutions, whose
instability has been proved analytically\cite{TV96}. Computer
simulations\cite{TT96} of an equivalent version of this model for
an order parameter $u(x,t)\equiv\int v(x,t)dx$ showed that even at
extremely small $\epsilon=10^{-4}$, the system exhibits STC.
However, this simulation did not provide any quantitative results
about the STC. The only result of this kind is in ref.\cite{KM97},
in which just a single quantitative characteristic, namely the
dependence of the mean amplitude of chaotic patterns on the control
parameter, was studied. In this Letter we span the gap in our
knowledge of this new type of STC, providing a detailed
quantitative description of its most important properties based on
numerical integration of Eq. (1). The simulations were carried out
using the pseudo-spectral method combined with a fourth-order
predictor-corrector integrator, for several different values of $L$
and two values of $\epsilon$ (0.2 and 0.5, respectively)\cite{NID}.
The Lyapunov exponents were calculated by linearizing the equation
along the trajectory, performing a re-orthonormalization after a
few integration steps to prevent the largest Lyapunov exponent from
swamping all the others\cite{PC89}. A typical pattern $v(x,t)$ as a
function of space and time in the steady chaotic regime is shown in
Fig.~1. The time averaged power spectrum $\langle |\hat
v(k)|^{2}\rangle $ obtained over a time period of $T=10^{4}$ for
system size $L=78$ with $\epsilon=0.2$ is shown in Fig.~2. As can
be seen, the dominant modes occur in the vicinity of $k=\pm 1$.
Note, however, the smaller peak near zero wavenumber, which arises
from coupling between unstable short-wavelength modes centered
about $k=\pm 1$ and the slowly decaying modes from the Goldstone
branch of the spectrum centered at $k = 0$ (using terminology based
upon the linear stability analysis of the spatially uniform
solution). The power spectrum for $\hat u(k)=\hat v(k)/ik$ (shown
in the insert in Fig.~2) has the dominant peak near $k=0$, as was
originally demonstrated in\cite{TT96}. One would also expect on
general grounds\cite{HS89} each Fourier transform variable $\hat
v(k,\omega)$ to be governed by a Gaussian probability distribution
functional, $\exp(-D(|\hat {v}(k,\omega)|^{2})$, for small $k$ and
$\omega$. We have verified this for several values of $k$ and
$\omega$. In particular, for $\omega=0.6$ the Gaussian distribution
holds for $0 < k < 2$ (see Fig. 3).

\begin{figure}
\epsfxsize = 0.45\textwidth
\epsfysize = 0.4\textwidth
\makebox{\epsfbox{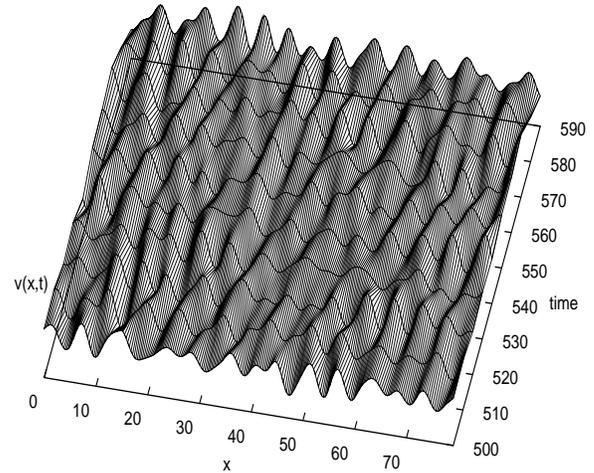}}
\caption{Typical space-time configuration for $v(x,t)$
in the steady chaotic regime is shown here. The configuration shown has
evolved from random initial condition within a system size $L=78$
and $\epsilon=0.5$.\label{fig1}}
\end{figure}

\begin{figure}
\epsfxsize = 0.45\textwidth
\epsfysize = 0.35\textwidth
\makebox{\epsfbox{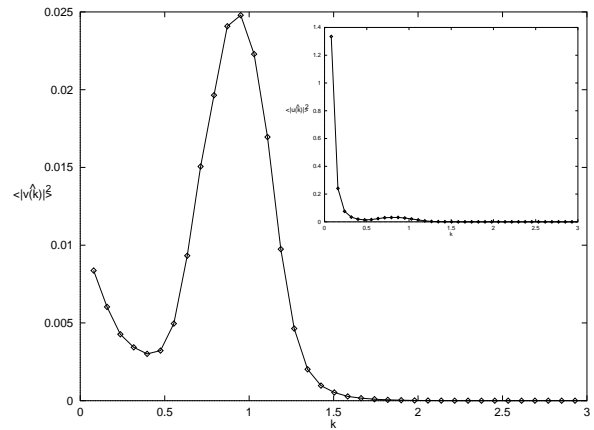}}
\caption{Time average power spectra $\langle|\hat {v}(k)|^{2}\rangle$
and $\langle|\hat {u}(k)|^{2}\rangle$ in $k$-space are plotted. The
system size is $L=78$ and control parameter $\epsilon=0.2$. Note
that the power spectrum has symmetry with respect to $ k\rightarrow
-k$.\label{fig2}}
\end{figure}

As has already been emphasized, the key question for any system
exhibiting STC is whether it can be represented as a union of
weakly correlated subsystems. If this is the case, then the
spectrum of Lyapunov exponents for the entire system should be {\it
intensive} in the sense that $\lambda_{i}$ is a function only of
the intensive index $i/V$, i.e., $\lambda_{i}=f(i/V)$, where V
stands for the volume of the system\cite{HG98}. The question is not
trivial for the type of STC considered here, because of the
importance of the long-wavelength modes and the divergence of the
two point correlation length $\xi_{2}$\cite{MT99} as $\epsilon
\rightarrow 0$\cite{MIT97}. To answer this question, a detailed
study of the Lyapunov spectrum for the Nikolaevskii model was
conducted. The results are shown in Fig. 4, where the number of
Lyapunov exponents greater than a particular value $\lambda_{i}$,
scaled by the system size $L$, is plotted versus $\lambda_{i}$ for
$\epsilon=0.5$. A similar curve is found at $\epsilon=0.2$, but the
maximum positive eigenvalue is now smaller (one expects this
eigenvalue to vanish as $\epsilon \rightarrow 0$.) The intensive
nature of the Lyapunov density is evident. In this case, one also
expects the fractal dimension $D(L)$ of an attractor to be
extensive for large enough $L$ (extensive chaos), as was first
shown by Manneville\cite{PM85} for chaotic solutions of the
Kuramoto-Sivashinsky equation. We have checked this for the
Nikolaevskii model using the Kaplan-Yorke formula\cite{KY79} for
the Lyapunov dimension,
\begin{equation}
D(L) = K + \sum_{i=1}^{K} \lambda_{i}/|\lambda_{k+1}|
\end{equation}
where the integer $K$ is the largest integer such that the sum of
the first $K$ Lyapunov exponents is nonnegative. We have also
calculated the Kolmogorov-Sinai entropy $H(L)$ from the definition
\begin{equation}
H(L) = \sum_{i=1}^{i_{+}} \lambda_{i}
\end{equation}
where the sum is over the positive Lyapunov exponents. The
Kolmogorov-Sinai entropy\cite{PC89} is a measure of the mean rate
of information production in a system, or the mean rate of growth
of uncertainty in a system subjected to small perturbations. We
find that for large enough $L$ both $D(L)$ and $H(L)$ are
extensive. Our results for the Lyapunov dimension $D(L)$ are shown
in Figure 5 for $\epsilon=0.5$. The same behavior is found at
$\epsilon=0.2$, but with a different slope (naturally the slope of
the straight line D(L) at $\epsilon=0.2$ is smaller than that at
$\epsilon=0.5$). In addition, we find that the upper index $i_{+}$
in (3), which corresponds to the smallest positive Lyapunov
exponent, is also proportional to $L$.

\begin{figure}
\epsfxsize = 0.45\textwidth
\epsfysize = 0.35\textwidth
\makebox{\epsfbox{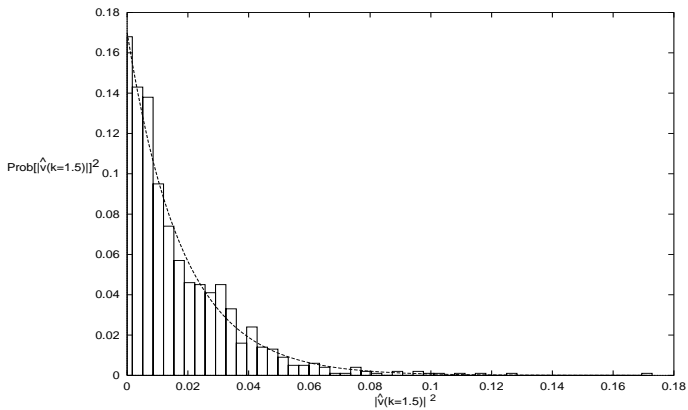}}
\caption{Probability density distribution of
power spectrum $\hat {v}(k)$ for $k=1.5$ with system size $L=78$
and $\epsilon=0.5$. Notice that the validity of the Gaussian
distribution for $\hat v(k)$ (see the main text) implies an
exponential distribution $f(z)=D{\rm e}^{-D z}$ for the variable
$z\equiv |\hat {v}(k)|^{2}$. In this case we find the value
$D\approx 47$.\label{fig3}}
\end{figure}

The important characteristic of STC is the dimension correlation
length $\xi_{\delta}$\cite{CH93,HG98}. This length is defined as
$\xi_{\delta} \equiv \delta^{-1/d}$, where $\delta \equiv
lim_{L\rightarrow \infty} D(L)/L^{d}$. It can be thought of as the
``radius" of a volume that contains one degree of freedom, or, as
the linear size of the subsystem described above. The value of this
dimension correlation length for Eq.~(1) is $\xi_{\delta}= 3.0$ for
$\epsilon=0.5$ and $\xi_{\delta}= 3.3$ for $\epsilon=0.2$. In
contrast to $\xi_{\delta}$ the above-mentioned two point
correlation length $\xi_2$ governs the spatial decay of the
correlation function. In general these two lengths are
different\cite{HG98}. We found that $\xi_{2} \cong 4.9$ for the
Nikolaevskii model at $\epsilon = 0.5$ and $\xi_{2}
\cong 5.6$ at $\epsilon=0.2$.

Finally, Tajima and Greenside\cite{TG99} have recently found for
the one dimensional Kuramoto-Sivashinsky model that $D(L)$ is also
"microextensive." Namely, they found that if one increases L by a
small amount $\delta L$, with $\delta L\ll \xi_{\delta}$ ($\delta
L=0.8$ in our simulation), one finds that $D(L)$ satisfies the same
linear relationship as that characterizing extensive chaos. We have
examined this for two different domains of $L$ for our model and
found that microextensivity holds for both $D(L)$ and $H(L)$. Our
results for microextensivity for $D(L)$ are shown in the insert in
Fig. 5. 

\begin{figure}
\epsfxsize = 0.45\textwidth
\epsfysize = 0.35\textwidth
\makebox{\epsfbox{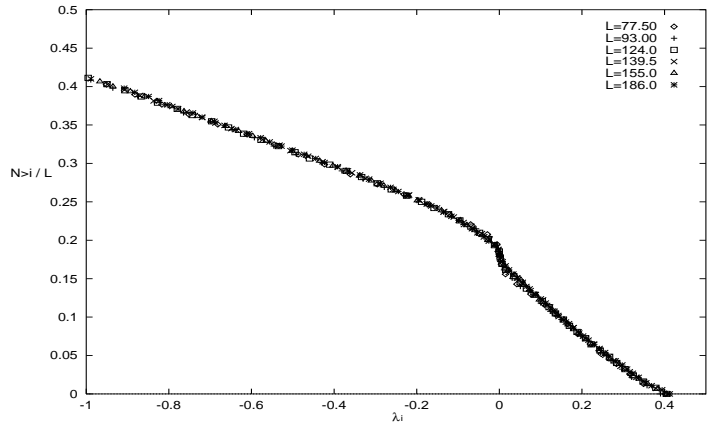}}
\caption{Here $N_{>i}$ is the number of Lyapunov exponents greater than
a particular value $\lambda_{i}$. We plot $N_{>i}/L$ (scaled by the
system size $L$) vs $\lambda_{i}$ in the case $\epsilon=0.5$.\label{fig4}}
\end{figure}

In conclusion, our detailed study of this new type of STC based on
the numerical integration of the Nikolaevskii model shows that for
sufficiently large system size the chaos is both extensive and
microextensive. We also found that the system satisfies Gaussian
statistics at sufficiently small wavenumbers and frequencies. We
believe these results are quite general and reflect intrinsic
features of this type of STC, rather than specific peculiarities of
the model.

There are several interesting questions to investigate in the limit
$\epsilon \rightarrow 0$, including the dependence of quantities
such as the correlation lengths and Lyapunov exponents on
$\epsilon$ as well as the possible scaling of the power spectrum.
This study is in progress and will be reported elsewhere.

\begin{figure}
\epsfxsize = 0.45\textwidth
\epsfysize = 0.35\textwidth
\makebox{\epsfbox{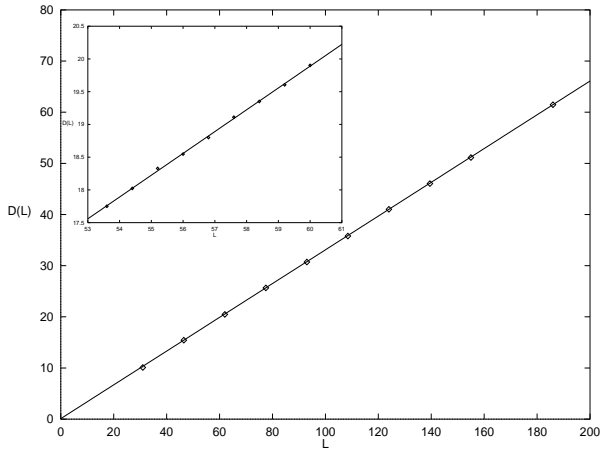}}
\caption{Lyapunov dimension $D(L)$ vs system size $L$ for $\epsilon=0.5$.
$\delta L=15.5$ for extensive, and $\delta L=0.8$ for
microextensive.\label{fig5}}
\end{figure}

We would like to thank Henry Greenside for a helpful discussion.
This work was supported by a grant from NATO CRG.CRG972822, by NSF
Grant DMR9810409, projects PB94-1167 and PB97-0141-C02-01 (Spain)
and the Grant-in-Aid for Scientific Research (No. 11837006) from
the Ministry of Education, Science, Sports and Culture (Japan). We
also wish to acknowledge an allocation of time on the Pittsburgh
Supercomputer Center, where some of this work was carried out.

\end{twocolumns}


\begin{references}
\bibitem{N} It is relevant to mention that the
consideration of convection with the stress-free boundary
conditions\cite{SZ82,BB84} was the very first analysis of pattern
formation in the degenerate systems. However, at that time the
dramatic destabilization of steady patterns obtained in this case
was regarded as a peculiarity of the particular problem,
rather than a generic attribute of the degeneracy.
\bibitem{NID} Although the original numerical study\cite{TT96}
was carried out in terms of the order parameter $u(x,t)$, we have
found that Eq.~(1) is more convenient for numerical study as it
exhibits greater numerical stability. The reason is that the
conservation law $d\bar v/dt=0$ holding for this equation (where
the bar denotes a spatial average) helps to stabilize the code. In
our simulations the value $\bar v=0$ was imposed by the initial,
random, conditions. For the space discretization we have found that
$\Delta x=0.31$ provides a good enough resolution.
\bibitem{MT99}At small $\epsilon$ in this model the two point
correlation function ($\langle v(x,t)v(x+x',t)\rangle$, where the
$\langle\ldots\rangle$ denotes the time average) is generally
oscillatory with a spatial period of the oscillations close to
$2\pi$ and a slow decay of its amplitude over a scale $\xi_{2}$.
Thus, the model has two correlation length scales, with $\xi_2$
being the larger of the two.
\end{references}
\end{document}